\documentstyle[12pt]{article} 
\begin{document}
\begin{titlepage}
\begin{center} 
\vskip 1.0cm
{\Large\bf Prime decomposition and correlation measure\\ 
of finite
quantum systems}
\vskip 1.0 cm

{\bf D. Ellinas$^1$  }\renewcommand{\thefootnote}{\diamond}\footnote{ Email:  {\tt
ellinas@aml.tuc.gr}}
and {\bf  E. G. Floratos$^2$}
\renewcommand{\thefootnote}{\triangleleft}\footnote{ Email:  {\tt
floratos@cyclades.nrcps.ariadne-t.gr}}

\vskip 0.2cm 
$^1$ Applied Mathematics and Computers Lab.\\
Technical University of Crete \\ 
GR - 73 100 Chania Crete Greece\\
\vskip 0.5cm
$^2$ NCRC Demokritos, Institute of Nuclear Physics\\
GR - 15 310 Ag. Paraskevi 
 Attiki Greece\\
and \\
Department of Physics, University of Crete, Crete Greece.

\vskip 2.0 cm 
{\bf Abstract} 
\end{center} 
\vskip 0.2 cm 
Under the name prime decomposition (pd), a unique decomposition of an 
arbitrary $N$-dimensional density matrix $\rho$
into a sum of seperable density matrices with dimensions determined by the coprime
factors of $N$ is introduced.  For a class of density matrices a complete
tensor product factorization is achieved. The construction is based on the 
Chinese Remainder Theorem, and the projective  unitary representation of $Z _N$ 
by the discrete Heisenberg group $H_N$. The pd isomorphism
 is unitarily implemented and 
it is shown to be coassociative and to act on $H_N$ as comultiplication.
Density matrices with complete pd are interpreted as grouplike elements of 
$H_N$.
To quantify the distance of $\rho$ from its pd a trace-norm correlation
index $\cal E$ is introduced and its invariance groups are determined.

\centerline{03.65 Bz, 42.50 Dv, 89.70.+c}
\centerline{Journal of Physics A: Math. Gen. {\bf 32} (1999) L63-L69}
\end{titlepage}

\font\blackboardss=msbm5 
%
%
\def\half{\scriptstyle{\frac{1}{2}}}
\def\halft{\textstyle{\frac{1}{2}}} 
\def\osqrt{\textstyle{\frac{1}{\sqrt2}}}
\def\lsqrt{\textstyle{\frac{\l}{\sqrt2}}} 
\def\phalf{\textstyle{\frac{\pi}{2}}}
\newcommand{\fscr}[2]{\scriptstyle \frac{#1}{#2}} 
\def\cc{\hbox{C\kern-0.55em\raise0.4ex\hbox{$\scriptstyle |$}}}
\def\zz{\hbox{{\sf Z}\kern-0.45em\raise0.0ex\hbox{\sf Z}}}     
\def\rr{\hbox{R\kern-0.55em\raise0.4ex\hbox{$\scriptstyle |$}}}  
\def\ena{1\hskip-0.25truecm 1}  
\def\zita{{\zz}^{ 2}_{N}}
\def\zitas{{\zz}^{* 2}_{N}}
\def\ccd{{\cc\hskip0.2truecm}^{2}}
\def\cct{{\cc\hskip0.2truecm}^{3}}                                          

\def\pf{{\it pd \/}}
\def\ie{{\it i.e \/}}
\def\tr{{\rm Tr \/}}
\def\eg{{\it e.g \/}} 
\def\cf{{\it c.f \/}} 
\def\viz{{\it viz. \/}}
\def\ad{a^\dagger} 
\def\ab{\bar{\alpha}} 
\def\nub{\bar{\nu}}
\def\hi{\chi_{klm}} 
\def\udp{U_{\lambda}^{\dagger}}
\def\udm{U_{-\lambda}^{\dagger}} 
\def\utp{\tilde{U}_{\lambda}}
\def\utm{\tilde{U}_{-\lambda}} 
\def\up{U_{\lambda}} 
\def\um{U_{-\lambda}}
\def\so{\sigma_{1}} 
\def\st{\sigma_{2}} 
\def\sth{\sigma_{3}}
\def\sp{\sigma^{+}} 
\def\sm{\sigma^{-}} 
\def\ovl{\overline}

%
\newcommand{\bra}[1]{\left<#1\right|} \newcommand{\ket}[1]{\left|#1\right>}
\newcommand{\braket}[1]{\left<#1\right>}
\newcommand{\inner}[2]{\left<#1|#2\right>}
\newcommand{\sand}[3]{\left<#1|#2|#3\right>}
\newcommand{\proj}[2]{\left|#1\left>\right<#2\right|} %
\newcommand{\rbra}[1]{\left(#1\right|} \newcommand{\rket}[1]{\left|#1\right)}
\newcommand{\rbraket}[1]{\left(#1\right)}
\newcommand{\rinner}[2]{\left(#1|#2\right)}
\newcommand{\rsand}[3]{\left(#1|#2|#3\right)}
\newcommand{\rproj}[2]{\left|#1\left)\right(#2\right|}
\newcommand{\absqr}[1]{{\left|#1\right|}^2}
\newcommand{\abs}[1]{\left|#1\right|}
\newcommand{\pl}[2]{\partial_{#1}^{#2}} 
\newcommand{\plz}[1]{\partial_{z}^{#1}}
\newcommand{\plzb}[1]{\partial_{\overline{z}}^{#1}} 
\newcommand{\zi}[1]{ z^{#1}}
\newcommand{\zib}[1]{{\overline{z}}^{#1}} 
\newcommand{\mat}[4]{\left(\begin{array}{cc} #1 & #2 \\ #3 & #4
\end{array}\right)} %
\newcommand{\col}[2]{\left( \begin{array}{c} #1 \\ #2
\end{array} \right)} 
\def\a{\alpha} 
\def\b{\beta} 
\def\g{\gamma}
\def\d{\delta} 
\def\e{\epsilon} 
\def\z{\zeta} 
\def\th{\theta} 
\def\f{\phi}
\def\la{\lambda} 
\def\m{\mu} 
\def\p{\pi} 
\def\om{\omega}  
\def\D{\Delta} 
\def\zb{\bar{z}} 
\newcommand{\lag}[2]{L_{#1}^{#2}(4\l^{2})}
\newcommand{\mes}[1]{d\mu(#1)} 
\def\nd{\noindent}
\def\nn{\nonumber} 
\def\cap{\caption} 
\def\cline{\centerline}
\newcommand{\be}{\begin{equation}} 
\newcommand{\ee}{\end{equation}}
\newcommand{\ba}{\begin{array}} 
\newcommand{\ea}{\end{array}}
\newcommand{\bea}{\begin{eqnarray}} 
\newcommand{\eea}{\end{eqnarray}}
\newcommand{\beann}{\begin{eqnarray*}} 
\newcommand{\eeann}{\end{eqnarray*}}
\newcommand{\bfg}{\begin{figure}} 
\newcommand{\efg}{\end{figure}}



Quantum correlations, an emblematic notion of quantum theory, remains an open
challenge since the early days of Quantum Mechanics~\cite{epr,b}.  Recent
investigations have set important questions concerning 
classification of various
types of quantum correlations and their appropriate quantification.  These
theoretical activities have parallel developments with, and are 
partly motivated from recent
interesting proposals which engage quantum correlations to such diverge tasks 
as
\eg quantum computation and communication~\cite{qcomp1,qcomp2}, 
quantum cryptography~\cite{ekert}, teleportation~\cite{tele}, 
and some new frequency standards~\cite{freq}.  
Although
the classification of quantum correlations is still open to refinements,
it appears to
include the following cases:  for pure states, correlations entail nonlocality
and give rise to violation of Bell inequalities~\cite{b}.  For mixed states, two
systems are considered uncorrelated if the composite system density matrix
factorizes into a product of reduced density matrices, one for 
each isolated quantum
subsystem \viz $\rho = \rho_1 \otimes \rho_2$, where 
$\rho_{1,2} = \tr_{1,2} \rho $, are determined by partial tracing. 
Quantification measures for that case include the von Neumann entropy~\cite{bp}
 and other invariant indices~\cite{sm}. On the other 
 hand classical correlations for quantum subsystems imply seperability of 
the joint system density matrix, which is analysed into a convex sum for
products of pure states \viz $\rho = \sum_{i} p_{i} 
\rho_{1}^{i}\otimes \rho_{2}^{i}$, $0\leq p_{i} \leq 1$, $\sum_{i} p_{i} = 1$,
~\cite{werner}.
Necessary and sufficient conditions for the existence of such convex 
decompositions for $\rho$'s acting on $\ccd \times \ccd$ 
and
 $\ccd \times \cct$, became available recently~\cite{p,horo}.
Upper bounds for the number of terms in such convex expansions
of seperable matrices have also been determined, 
together with construction algorithms 
for the cases dim${\cal H} \leq 6 $~\cite{san} and 
dim$\cal H \leq \infty$~\cite{ls}. Beyond these types of classical
correlations we encounter inseperable or entangle quantum states. For 
their characterization and  the quantification of their entanglement 
some general
conditions have been presented that good entanglement measures should 
satisfy~\cite{vedral}.

In this Letter we address the problem of the 
$\it prime \; decomposition \; (pd)$,
of a finite but otherwise arbitrary $N$-dimensional square density matrix $\rho$, into a sum of 
products of elementary density matrices,
the number and 
the respective dimensions of which are determined by the compositeness of the
dimension of $\rho$.
This is achieved by means of  $\it 1)$ the so called Chinese 
Remainder Theorem (CRT)~\cite{knuth},
that is based on the prime decomposition of $N$ (this also explains 
the name 
we have chosen for the decomposition), and $\it 2)$ by the 
fact that 
the discrete Heisenberg group $H_N$, provides a projective 
representation of the abelian cyclic group $\zz_N$~\cite{swin}. More concretely,
if $N=p_{1}^{m_1}p_{2}^{m_2}\ldots p_{t}^{m_t}$, is the prime factor 
decomposition of $N$, where $p$'s are distinct primes, then the 
\pf of the
density matrix
involves square matrices $\rho^{(i)}$ $i=1,\ldots, t$, with power prime dimension
equal to $N_{i} = p_{i}^{m_i}$. Also the number  $t$ of $\rho$-factors 
 is bounded by the number of coprime factors of $N$.
As a measure of the correlation of a given mixed state $\rho$, 
with its possible prime or other decomposition, we evaluate the 
trace-norm distance between the two densities, study its unitary invariant
symmetries, and interpret it in terms of the quantum variances between 
local operators of the subsystems of the decomposition. 

We start by considering the matrix realization of the discrete 
Heisenberg group $H_N$ generated by the operator set of $N^2$ elements
 $J_m \equiv J_{m_1 m_2} = \omega^{\frac{1}{2}{m_1 m_2}} g^{m_1} h^{m_2}$, where the matrices
\bea 
\label{gandh} 
g &=& {\rm diag}(1,\omega,\ldots,\omega^{N-1}), \nn \\ 
h &=&
\sum_{n\in{\bf Z}_{N}}\ket{n}\bra{n+1}\;, 
\eea 

\nd satisfy the relations $\omega gh=hg$,
$h^{\dagger}=h^{-1}$, $g^{N}=h^{N}=\ena $, $hh^{\dagger}=h^{\dagger}h=\ena $,
$gg^{\dagger}=g^{\dagger}g=\ena $,  with $\omega ^N=1$, and
$(m_1,m_2)\in \zita $,
the square index-lattice. By virtue of these
relations the following
commutators are valid ~\cite{balian},~\cite{zachos}, ~\cite{manol}: 
\be
\left[J_{m},J_{n}\right]=-2i \sin\left[\frac{\pi}{N} m\times n\right] J_{m+n} \;
\pmod{N}\;.
\ee 

\nd Moreover due to linear independence, completeness and 
the orthonormality issued by the relation 
\be 
{\rm Tr} J_m J_n = N
\delta_{m+n,0}\; \pmod{N}\;,
\ee 

\nd the same generator set forms a basis of the $su(N)$ matrices~\cite{swin}.

Let us consider a $N$-dimensional quantum system $\cal S$ with Hilbert space
${\cal H}_N$. The generators of the finite Heisenberg
group $H_N$ provide an operator basis $\{ J_{m} | m\in \zz_{N}^{2}\}$,
for the decomposition of the density matrix $\rho$ of  $\cal S$, \viz
\begin{equation}
 \label{rom} 
 \rho=\frac{1}{N}
\sum_{m\in \zita} (\lambda_{m}
J_{m}) =
 \frac{1}{N}[
{\rm \ena} + \sum_{m\in \zitas} \lambda_{m}
J_{m}]\;. 
\end{equation} 

\nd with  $\zitas\equiv \zz_{N}\times \zz_{N} \setminus (0,0)$, 
We note here that due to the Hermitian conjugation of the basis
elements \ie $J^{\dagger}_{m}=J_{-m_1 , -m_2 }=J_{N-m_1 ,
N-m_2 }$, the Hermiticity of the density matrix implies for its elements
the reality conditions $\lambda_{m}^{*}=\lambda_{N-m}$. Let us assume 
 $N$ to be a composite positive integer with prime-power decomposition
$N=p_{1}^{n_1 }p_{2}^{n_2 }\ldots p_{t}^{n_t }\equiv N_1 N_2 \ldots N_t$, 
where each 
of the factors is distinct, uniquely determined and 
relatively prime to each other, \ie $\gcd(N_i , N_j ) =1$ when $i\neq j$.
Then according to the CRT the isomorphism 
$\zz_N \cong \zz_{N_1 } \oplus \cdots \oplus \zz_{N_t }$, is valid for the index-
cyclic groups labelling the operator basis. To proceed we introduce the 
group isomorphic map $\zita \stackrel{\d}{\rightarrow} \zz_{N_1 }^{2}
 \oplus \cdots \oplus \zz_{N_t }^{2}$, between the cyclic groups.
 The explicit definition reads: $(m_1 ,m_2 ) \stackrel{\d}{\rightarrow}
 (\d (m_1),\d(m_2))\stackrel{\d}{\rightarrow}
 (m_{11} ; m_{21} , m_{12} ; m_{22},\ldots , m_{1t} ; m_{2t}  )$,
 where $m_{1i} = m_1 -p N_{i} $, $m_{2i} = m_2 - q N_{i} $, $i=1, \ldots, t$, 
 $p, q \in \zz$, stand for the residues of the division of $m_1, m_2$ by $N_i$.
 
 Next we regard the fact that $H_N$, provides a projective unitary
 representation
 of the additive cyclic group $\zita$, by means of the map 
 $\zita \stackrel{\pi_{N}}{\rightarrow} H_N $. More explicitly,
 $(m_1 ,m_2) \stackrel{\pi_{N}}{\rightarrow} \pi_{N} (m_1 , m_2) = 
 J_{m_1,m_2 }$, with
 the property $\pi_{N}(m+n)=J_{m+n}=J_{m}J_{n}e^{\frac{i}{2}m\times n}=
 \pi_{N}(m)\pi_{N}(n) e^{\frac{i}{2}m\times n}$, where $m\times n :=
 m_{1} n_{2} - m_{2} n_{1}$.
Then the following commuting diagram:
\vspace{0.3cm}
\[
\begin{array}{rcccl}
 & \zz ^2_N & \stackrel{\d}{\longrightarrow} & 
 \zz ^ 2_{N_1 }\oplus \cdots \oplus \zz ^2_{N_t } & \\
 \pi _N & \downarrow & & 
\downarrow &  \pi _{N_1 }
\times \cdots \times  \pi _{N_t } \\
 & H_N & 
 \begin{array}[t]{c}\longrightarrow \\ \pi _{\d} \end{array}& 
H_{N_1 }\otimes  \cdots \otimes H_{N_t } & 
\end{array} 
\]
\vspace{0.2cm}
\centerline{Fig. 1}

\nd given by the equation $\pi _{\d} \circ \pi _{N} = (\pi _{N_1 }
\times \cdots \times \pi _{N_t }) \circ \d $,
induces the isomorphism of CRT from the index-groups 
to the associated Heisenberg groups by
the following component version of the above diagram:

\vspace{0.3cm}
\[
\begin{array}{rcccl}
 & m & \stackrel{\d}{\longrightarrow} & \d(m)= 
 (m_{11} ; m_{21} , \ldots , m_{1t} ; m_{2t}  ) & \\
\pi _{N} 
& \downarrow & & 
\downarrow & 
\pi _{N_1 }
\times \cdots \times 
\pi _{N_t } \\
 & \pi _{N}(m) = J_{m} & 
 \begin{array}[t]{c}\longrightarrow \\ \pi _{\d}   
\end{array}& 
\pi _{\d} (J_{m}) = J_{\d (m)} =
  J_{(m_{11} ; m_{21} , \ldots , m_{1t} ; m_{2t}  ) }
 &
\end{array} 
\]
\vspace{0.2cm}
\centerline{Fig. 2}

We state the main proposition for the prime decomposition:

{\it Proposition.} The isomorphism $\pi_{\d}$, determined by the commuting
diagram of Fig. 1, via its component version Fig. 2, is a linear
map which induces the $\d$-map of CRT into the Heisenberg group
$H_N$, and provides the unique \pf of elements of $H_N$. Also $\pi_{\d}$ is 
implemented by unitary operator in the Hilbert space ${\cal H}_N$ and possesses
the coassociativity property.

\nd {\it Proof.} If $m\in \zz ^2_N$ and $J_m = \omega^{1/2 m_1 m_2} g^{m_1} 
h^{m_2}$, then
$\d(m_1 ,m_2 )=  
(m_{11} ; m_{21} ,\ldots , m_{1t} ; m_{2t}  ) $, with
$m_{1i}$ and $m_{2i}$, the residues of the division  of $m_1 , m_2$  by
$N_i$ respectively. According to CRT $\d$ is an isomorphism
the determination of which provides the solution of a system of
congruences $m_1 \equiv m_{1i} \pmod{N_{i}}$ and $m_2 \equiv m_{2i} 
\pmod{N_{i}}$,
when $\gcd (N_{i}, N_{j})=1, N_{i}\neq N_{j}$, \ie when
$N_{i}, N_{j} \ \ \ i=1,\ldots, t$,
are pairwise coprime positive integers. 
Inversely, given the residues, the numbers $m_1 ,m_2$ can be determined in a
mixed-radix notation by $m_1 \equiv \sum_{i=1}^{t}m_{1i}\ovl{N}_{i} y_{i}$
and  $m_2 \equiv \sum_{i=1}^{t}m_{2i}\ovl{N}_{i} y_{i}, \pmod{N}$, where
$\ovl{N}_{i} := \frac{N}{N_i }$ and $ y_{i}$ is the solution of the
congruence $\ovl{N} _{i} y_{i} \equiv 1, \pmod{N_i} $. Alternatively
by means of the Euler function $\phi(k)$, which counts the positive
integers $l\leq k$, which are coprime to $k$, the $y_i$
is given by $y_{i} \equiv \ovl{N}_{i}^{\phi (N_i) - 1} , \pmod{N_i }$.
Then $m_1 ,m_2$, are expressed in the form $m_1 \equiv \sum_{i=1}^{t}
m_{1i}  \ovl{N}_{i}^{\phi (N_i )}$ and  
$m_2 \equiv \sum_{i=1}^{t}
m_{2i}  \ovl{N}_{i}^{\phi (N_i )} \pmod{N}$.

We turn now to study the consequences of this decomposition for the
generators of $H_N$. With the notation as before 
we obtain the relations
\be
g^{m_1 }=g^{\sum_{i=1}^{t} m_{1i}  \ovl{N}_{i}^{\phi (N_i )}}=
\prod_{i=1}^{t}g_{i}^{m_{1i}}
\ee

\nd where $g_{i} := g^{\ovl{N}_{i}\phi (N_i ) }$ and
$g_{i}^{N_i}=g^{\ovl{N}_{i} N_{i} \phi (N_i )}=g^{N \phi (N_i ) }=
\ena $. Analogous relations
hold for the generator $h^{m_2}$. By direct computations it is verified that
$g_{i} h_{j} = h_{j}g_{i}$ if $i\neq j$, and $g_{i}^{k}h_{i}^{l} =
\omega ^{kl}_{i} h_{i}^{l}g_{i}^{k}$, where
$\omega _i := \omega^{\ovl{N} _{i}^{2 \phi (N_i )}} $.
This definition implies that $\omega_i$ is periodic with respect
to the coprime factors of $N$ \ie $\omega _{i}^{N_i} =
\omega ^{N_{i}\ovl{N} _{i}^{2 \phi(N_i )}}=1$,  for $i= 1, \ldots , t$.
Using the above commutation properties of the generators we write:
\bea
\pi_{\d} \circ \pi_ {N}(m_1 ,m_2 ) &=& \pi_{\d} (J_{m_1 m_2})=
 \prod_{i=1}^{t}
\omega _{i}^{1/2 m_{1i} m_{2i}} g_{i}^{m_{1i}} h_{i}^{m_{2i}} \equiv
\prod_{i=1}^{t} J_{m_{1i} m_{2i}}^{(i)} \nn \\ 
& \cong & \otimes_{i=1}^{t} J_{m_{1i} m_{2i}}^{(i)} = (\pi_{N_1} \times \cdots
\pi_{N_t}) (m_{11} m_{21} , \ldots , m_{1t} m_{2t}).
\eea

\nd The isomorphism introduced above is based on the fact the the
$J_{m_{1i} m_{2i}}^{(i)}$'s are commuting for different $i$'s and their
moduli  make them to behave as copies ($\pi_{\d}$-isomorphic images )
of the original
$J_{m_1 m_2 } \in H_N$, with periodicities $N_i \leq N$; this is 
similar to harmonics in Fourier analysis. The following embedding provides the explicit
form of the isomorphism:
\bea
J_{m_{1i} m_{2i}}^{(i)} &\cong& \ena_{N_1} \otimes \cdots
\otimes J_{m_{1i} m_{2i}} \otimes \cdots \otimes \ena_{N_t} 
= \pi_{N_i} (m_{1i} , m_{2i}) \in H_{N_i} ,
\eea

\nd with $m_{1i} , m_{2i} \in \zz ^2_{N_i}$; this provides
 the commutativity of the diagrams.

The prime decomposition of a general density matrix given in eq.~(\ref{rom})
with coefficients
$\rho_{m}={\rm Tr}_{N}(J_{m} \rho)$,
can now be evaluated and reads,
\bea
\pi_{\d}(\rho) &=& \frac{1}{N}\sum_{(m_1 ,m_2 )\in \zita} \rho_{\d(m_1 ,m_2)}
\pi_{\d}(J_{m_1 ,m_2}) =
\frac{1}{N}\sum_{(m_1 ,m_2) \in \zita} \rho_{\d(m_1 ,m_2)}
J_{\d(m_1 ,m_2 )} 
\nn \\
&\cong & \frac{1}{N}\sum_{(m_{11} m_{21}) \in \zz_{N_1}^2} \cdots
         \sum_{(m_{1t} m_{2t} )\in \zz_{N_t}^2}
\rho_{m_{11} m_{21} , \ldots , m_{1t} m_{2t} } J_{m_{11} m_{21}} \otimes \cdots
\otimes J_{m_{1t} m_{2t}}\;,
\eea

\nd where $\rho_{m_{11} m_{21} , \ldots , m_{1t} m_{2t} }  =
{\rm Tr}_{N_1} \cdots {\rm Tr }_{N_t}  (\pi_{\d}(\rho) 
J_{m_{11} m_{21}} \otimes \cdots
\otimes J_{m_{1t} m_{2t}})$.


\nd This suggests that
we first map $\rho$ into $ \pi_{\d}(\rho)$,
according to the previous analysis and
then project along the $J_{m_{1i}\; m_{2i}}^{(i)} $'s ,
in order to determine the coefficients of
the $\rho$-matrix factors in 
the prime decomposition.

A special form of \pf that contains only a single product term is possible for a
special class of  
density matrices with coefficients 
$\rho_m = \frac{1}{N} \omega ^{f(m)}$, where $f(m)\in l_2 ( \zita)$, an arbitrary
real function. If $ f(m)= \sum_{kl}f_{kl} m_1^k m_2^l $,  the tranformed 
coefficients $\rho _{\d (m)} = \frac{1}{N} \omega ^{f(\d (m))}$,
due to $\omega ^N =1$, factorize
 as follows:
\bea
\rho _{\d (m)} &=& \frac{1}{N} \omega ^{
\sum_{i=1}^{t} \sum_{kl} f_{kl} 
 m_{1i}^k m_{2j}^l \ovl{N}_i^{\phi (N_i)} \ovl{N}_j^{\phi (N_j)} 
 }\nn \\
 &=& \frac{1}{N} \omega ^{
 \sum_{i=1}^{t} \sum_{kl} f_{kl} 
 m_{1i}^k m_{2i}^l \ovl{N}_i^{2\phi (N_i)} 
 }\nn \\
 &=& \prod _{i=1}^{t} \frac{1}{N_i}\omega _i^{f(m_{1i}, m_{2i})} 
 =: \prod_{i=1}^{t} \rho _{(m_{1i}, m_{2i})} \;.
 \eea
 
 \nd Note that power raising is counted $\pmod{N}$, so above a Frobenious 
 type of map has been used \ie $m_1^k = (\sum_{i=1}^t m_{1i} \ovl{N}_i^{\phi (N_i) })^k =
 \sum_{i=1}^t m_{1i}^k \ovl{N}_i^{\phi (N_i)} \pmod{N} $. Therefore
 we obtain  the \pf $\pi_{\d} (\rho)=\rho ^{(1)} \otimes \cdots \otimes 
 \rho ^{(t)} $,
 where $\rho ^{(l)} = \sum_{(m_{1l}, m_{2l}) \in \zz_{N_l}^2} 
 \rho _{(m_{1l}, m_{2l})} J_{m_{1l} m_{2l}}$. In view of the coassociative
property of the $\pi_{\d}$, to be established shortly, we see that those matrices 
that admit such complete factorization behave as grouplike elements under the \pf map.  
 
 To proceed we study the unitary implementation of the \pf of density
 matrices. We introduce the operator
 $V_{\d} :{\cal H}_N \longrightarrow \otimes_{i=1}^t {\cal H}_{N_i} $,
 given by 
 \be
 V_{\d} =\sum_{n \in \zz _N} \proj{\d (n)}{n} \equiv 
\sum_{n \in \zz _N} \proj{\{n_i \}}{n} = \sum_{n \in \zz _N} 
\ket{n_1}\otimes \cdots  \otimes \ket{n_t}\bra{n} \;,
\ee

\nd and its conjugate
\be
V^{\dagger}_{\d} = V_{\d ^{-1}} = \sum_{\{n_i\}\in \{\zz _{N_i} \}}
\proj{\d ^{-1} (\{n_i\} ) }{\{n_i\} } \equiv \sum_{\{n_i\}\in \{\zz _{N_i} \}}
\ket{\d ^{-1} (n_1 , \ldots , n_t ) }\bra{n_1}\otimes \cdots 
\otimes \bra{n_t} \;,
\ee

\nd (this $\d$ as the one used earlier maps numbers to
their respective residues; see below).
These operators form a conjugate pair that obeys  the unitarity condition 
$V_{\d}V^{\dagger}_{\d} = \otimes_{i=1}^{t} \ena _{{\cal H}_{N_i}}$ and 
$V^{\dagger}_{\d}V_{\d}= \ena _{{\cal H}_{N}}$. Then it is straightforward to
verify that the \pf map $\pi_{\d}$ acting on general density matrix
 is implemented by the unitary similarity transformation
\ie $\pi_{\d}(\rho) = V_{\d}\rho V_{\d}^{\dagger}
\in \otimes_{i=1}^{t} H_{N_i}$. 

Finally we study briefly an important property of the \pf map $\pi_{\d}$,
namely that it becomes a coassociative comultiplication of the Heisenberg group
 $H_N$; this illustrates a connection of CRT with Hopf algebras~\cite{abe}
  in the framework
 of quantum mechanical correlations. 
Consider the map $\d_{n_1 , n_2} (x) =(x-\rho n_1 , x-\sigma n_2 )$, 
 $\rho , \sigma \in \zz $, by which a  $x\in \zz_{n_1 n_2}$, decomposes into its resedues 
 wrt coprimes $n_1 , n_2$. Also consider its dual map $\mu _{n_1 ,n_2} (a,b)
 \equiv an_2^{\phi (n_1)} + b n_1^{\phi (n_2)} =x \pmod{n_1 n_2}$, which
 constructs the solution of the congruences $x\equiv a \pmod{ n_1}$,  
 $x\equiv b \pmod{ n_2}$, according to CRT.  Then we check that
 for  $N_1 , N_2 , N_3 $ three coprime factors of $N$, the following
 equation is valid on any $a\in \zz _N$:
 \be
 (\d _{N_1 , N_2} \times id )\circ \d_{N_1 N_2 , N_3} = 
 (id \times \d_{N_2 , N_3} )\circ \d_{N_1 , N_2 N_3}\;.
 \ee
  
  \nd This is dual to the relation 
  \be
  \mu_{N_1 N_2 , N_3}\circ (\mu_{N_1 , N_2} \times id) =
  \mu_{N_1 , N_2 N_3} \circ ( id \times \mu_{N_2 , N_3} ) \;,
  \ee
  
  \nd which holds if we are given  three congruences and  combine them 
  pairwise in two different ways. This (co)associativity of the CRT maps, 
  in turn is induced into the \pf map $\pi_{\d}$, where it takes the form
  \be
  (\pi_{\d _{N_1 , N_2}} \otimes id )\circ \pi_{\d_{N_1 N_2 , N_3}} = 
  (id \otimes \pi_{\d_{N_2 , N_3} })\circ \pi_{\d_{N_1 , N_2 N_3}}\;.
 \ee
 
\nd As an example we take the system
\bea
x&\equiv&2 \pmod{3} \;,\nn \\
x&\equiv&2 \pmod{4} \;,\nn \\
x&\equiv&3 \pmod{5} \;,
\eea

\nd with solution $x=38 \pmod{60}$, and obtain 
 \bea
  \mu_{3\cdot 4 ,5}\circ (\mu_{3 ,4} \times id) (2,2,3) &=& \mu_{3\cdot 4 ,5} 
  (2,3) = 38 \nn \\
  \mu_{3, 4\cdot 5} \circ ( id \times \mu_{4 , 5} ) (2,2,3) &=& 
  \mu_{3, 4\cdot 5} (2,18) = 38 \;.
  \eea
   
\nd Dualizing we recover the relation for the $\d$'s which induces the 
coassociativity of the \pf mapping:
  \be
  (id \otimes \pi_{\d_{4,5} })\circ \pi_{\d_{3, 4\cdot 5}} (\rho ) =
   (\pi_{\d _{3,4}} \otimes id )\circ \pi_{\d_{3\cdot 4 , 5}} (\rho ) \;,
 \ee
 
\nd for $\rho \in H_{60}$. Closing this proof we note that the integral 
$\int _N : H_N \longrightarrow \cc \/$, with definition
$\int _N \rho := \tr _N \rho $, is 
invariant under the comultiplication $\pi _{\d _{N_1 , N_2}}$, in the sense 
that $(\int _{N_1} \otimes \int _{N_2} )\circ \pi _{\d _{N_1 , N_2}} (\rho )=
\int _N \rho $ $\Box$

We turn now to the study of the correlation between finite quantum systems.
We start with two systems with state vector Hilbert spaces of dimension $N_1,
N_2$ respectively.  Any observable and density matrix is expressed by the
elements of the Lie algebra $u(N_1)$, $u(N_2)$ correspondingly.  For the density
matrix of system-1 \eg, 
\begin{equation}
\label{romcomp}
\rho^{(1)}=\frac{1}{N_1}[
{\ena}^{(1)} + \sum_{m\in \zz _{N_1}^{2 *}} \lambda_{m}^{(1)}
J_{m}^{(1)}]\;, 
\end{equation} 

\nd and similarly for system-2.  The choice of the
operator basis $({\ena }^{(i)},J_{m}^{(i)}), (i=1,2)$, for the Lie algebra
$u(N_i)\approx u(1)\oplus su(N_i)$ is an important one.
For a
composite system the density matrix reads~\cite{dembost}
\begin{eqnarray} 
\rho&=&\frac{1}{N_1 N_2}
 \Big[ 
 {\ena 
}^{(1)}\otimes{\ena }^{(2)} + \sum_{m\in \zz _{N_1}^{2 *} }
\lambda_{m}^{(1)} J_{m}^{(1)}\otimes {\ena }^{(2)} 
+
\sum_{n\in \zz _{N_2}^{2 *}} \lambda_{n}^{(2)} {\ena
}^{(1)}\otimes J_{n}^{(2)} \nonumber \\ &+& \sum_{m\in \zz _{N_1}^{2 *}
} \sum_{n\in \zz _{N_2}^{2 *}} \lambda_{mn}^{(1,2)}
J_{m}^{(1)} \otimes J_{n}^{(2)} \Big]\; ,
\end{eqnarray}

\nd where $\lambda_{m}^{(1)}\equiv \braket{J_{m}^{(1)}}={\rm Tr}(\rho \cdot
J_{m}^{(1)} \otimes {\ena }^{(2)})$, $\lambda_{m}^{(2)}
\equiv \braket{J_{m }^{(2)}}={\rm Tr}(\rho \cdot
{\ena }^{(1)} \otimes J_{m}^{(2)} )$ and $\lambda_{mn}^{(1,2)}\equiv 
\braket{J_{m}^{(1)}\otimes J_{n}^{(2)}}={\rm Tr}(\rho
\cdot J_{m}^{(1)} \otimes J_{n}^{(2)})$, the correlation components.  Also by
partial tracing we define $\rho^{(1)}={\rm Tr}_{2}\rho$, $\rho^{(2)}={\rm
Tr}_{1}\rho$.  To proceed with the definition of the correlation index we 
view the space of matrices $\rho \in u(N_1)\otimes u(N_2)\equiv {\cal G}$,
as a
norm space with Hilbert-Schmidt (HS) norm,
\begin{equation}
||A||_{(2)}\equiv
\sqrt{<A,A>}=({\rm Tr} A^{\dagger} A)^{1/2} =
\sqrt{\sum_{ij=1}^{N^2}|a_{ij}|^2}\;,
\end{equation}

\nd for $A=(a_{ij})\in {\cal
G}$.  This is essentially a Frobenius type matrix norm, which
 is unitarily invariant \ie $||UAY||=||A||$,
for $U,Y$ unitary (the lower index of the norm will be omitted hereafter).
  Then we propose the following

{\it Definition.}\ \ The correlation scalar index of two coupled finite
quantum systems been in a mixed state $\rho$ is defined as~\cite{dembost}
\begin{equation}
\label{edef}
{\cal E}\equiv ||\Delta\rho||^{2}=
||\rho -
\rho^{(1)}\otimes\rho^{(2)}||^{2}\;.
\end{equation}

\nd Index $\cal E$ provides us with a measure of correlation between the
coupled systems in terms of the difference of the factorized partial density
matrices from the density of the composite system. It is cast in the
form
\begin{eqnarray}
{\cal
E}&=&||\rho||^{2} - 2 {\rm Tr}(\rho \cdot
\rho^{(1)}\otimes\rho^{(2)}) + ||\rho^{(1)}||^{2}
||\rho^{(2)}||^{2} \nn \\
&=& \frac{1}{N_1 N_2}\sum_{m \in \zz _{N_1}^{2 *}
} \sum_{n\in \zz _{N_2}^{2 *}}
\left[\lambda_{mn}^{(1,2)}-
\lambda_{m}^{(1)}\lambda_{n}^{(2)}\right] \left[\lambda_{N_{1} - m, N_{2} -
n}^{(1,2)}-\lambda_{N_{1} - m}^{(1)} \lambda_{N_{2} - n}^{(2)} \right]\;,
\end{eqnarray}

\nd where $N_1, N_2$-modulo arithmetic
applies in the respective indices.

\nd The index $\cal E$ vanishes for product states and by 
 using the reality conditions of the $\lambda _i$'s \ie 
$\lambda ^{(\nu )*}_{m}=
\lambda^{(\nu )}_{m - N_{\nu }}$, $\nu = 1,2$, $\lambda ^{(1,2) *}_{m,n}=
\lambda^{(1,2)}_{N_1 -m , N_2 -n }$
we introduce the matrix $\Lambda _{mn} :=
\lambda_{mn}^{(1,2)}-
\lambda_{m}^{(1)}\lambda_{n}^{(2)}=\braket{J_{m}^{(1)}\otimes J_{n}^{(2)}}-
\braket{J_{m}^{(1)}}\braket{J_{n}^{(2)}}$, and re-express the index in the form
\be
{\cal E}= \frac{1}{N_1 N_2}\tr \Lambda \Lambda ^{\dagger} \;.
\ee

\nd This last expression suggests first, that the index $\cal E$ is 
determined by the trace of the covariance matrix of local observables
$J_{m}^{(1)}$ and $J_{n}^{(2)}$, and second that it is 
invariant
under general unitary  tranformations of the group $U(N_1 \cdot N_2 )$,
\ie $\Lambda \rightarrow {\cal U}^{\dagger} \Lambda {\cal U} \ \ ;
\ \ \Lambda ^{\dagger} \rightarrow {\cal U}^{\dagger} \Lambda ^{\dagger}
{\cal U}$, with ${\cal U} \in U(N_1 \cdot N_2 ) \subset U(N_1 )\otimes U(N_2 )$.
The last inclusion describes the fact that the invariance unitary group
of $\cal E$, is in general larger than the local unitary transformations
in which case the symmetry group factorizes (\cf~ \cite{sm}).

\nd Extensions to three and more coupled systems is
straightforward.  For three systems \eg the composite density matrix involves
terms of the operator basis where the $J_m$'s are embedded in all possible ways
in the 3-tensor space.  Also for the reduced matrices there are various
possibilities in this case \ie $\rho^{i,j}={\rm Tr}_{k}\rho$ and $\rho^{i}={\rm
Tr}_{jk}\rho$, with cyclic permutations of $(i,j,k)=(1,2,3)$.  This gives rise
to different correlation indices \ie
\begin{eqnarray}
{\cal E}_{123}&=&||\rho -
\rho^{(1)}\otimes\rho^{(2)}\otimes\rho^{(3)}||^{2}_{(2)} \;, \nn \\
{\cal
E}_{1(23)}&=&||\rho - \rho^{(1)}\otimes\rho^{(23)}||^{2}_{(2)} \;, \nn \\
{\cal
E}_{2(13)}&=&||\rho - \rho^{(2)}\otimes\rho^{(13)}||^{2}_{(2)} \;, \nn \\
{\cal
E}_{3(12)}&=&||\rho - \rho^{(3)}\otimes\rho^{(12)}||^{2}_{(2)} \;.  
\end{eqnarray}

Closing we should mention that the correlation index can be expressed in 
terms of the $P$ function of the involved density matrices, associated
with the $SU(2)$ group coherent state of dimension $N$. This possibility as 
will be explained elsewhere~\cite{megalo}, is based on the fact the
the $su(N)$ algebra generators used here
in the expansion of the $N$-dim density matrices, can be embedded 
(by means of the polar decomposition of the $su(2)$ algebra), into the 
enveloping algebra $U(su(2))$. Examples of the finite case together 
with extensions to
infinite dimensional quantum systems will also be reported elsewhere.

\vskip 1.0 cm
We acknowledge support from the Greek Secretariat of Research and Technology
under contract $\Pi ENE\Delta$ 95/1981.

\vskip 1.0cm

Figure captions.

Fig. 1. Induction of CRT into the Heisenberg group.

Fig. 2. Component version of the induction of fig.1.


\begin{thebibliography}{25}
\bibitem{epr}
A. Einstein,  B. Podolsky and N. Rosen,  Phys.  Rev.  {\bf 47} (1935) 777.
\bibitem{b}J. Bell, Physics (N.Y) {\bf 1}, (1964) 195; 

J. F. Clauser, Horne and R. A. Holt, Phys.  Rev.  Lett.  {\bf 23}, (1969) 880.  
\bibitem{qcomp1} D. Deutsch, Proc.  R.  Soc.  Lond.  Ser. A, {\bf 425},(1989) 73.  
\bibitem{qcomp2} C. H. Bennett et. al, Phys. Rev. A {\bf 53}, 2046 (1996) ;

D. Deutsch et. al, Phys. Rev. Lett. {\bf 77}, 28 (1996).
\bibitem{ekert} A. K. Ekert, Phys.  Rev. Lett.  {\bf 67}, (1991) 661. 
\bibitem{tele} C. H. Bennett et. al , Phys. Rev. Lett. {bf 67}, 661 (1991).
\bibitem{freq} S. Huelga, et. al Phys. Rev. Lett. {\bf 79}, 3865 (1997).
\bibitem{bp}S. M. Barnett and S. J. D. Phoenix, Phys.  Rev.  A {\bf 44}, (1991) 535.        
\bibitem{sm} J. Schlienz  and  G. Mahler, Phys.  Rev.  A {\bf 52},(1995) 4396.  
\bibitem{werner} R. F. Werner, Phys. Rev. A {\bf 40}, 4277 (1989).
\bibitem{p} A. Peres, Phys. Rev. Lett. {\bf 77}, 1413 (1996).
\bibitem{horo} M. Horodecki et. al Phys. Lett. A {\bf 223}, 1 (1996).
\bibitem{san} A. Sanpera et. al, e-print archive quant-ph/9703004 (1997).
\bibitem{ls} M. Lewenstein and A. Sanpera, e-print archive quant-ph/9707043 .
\bibitem{vedral} V. Vedral et. al, Phys. Rev. Lett. {\bf 78} 2275 (1997).
\bibitem{knuth} D. E. Knuth, Seminumerical Algorithms Vol. II, (Addison-
Wesley, Reading, Massachusetts, 1981).
\bibitem{swin} H. Weyl, The theory of groups and Quantum Mechanics, 
(Dover, New York, 1950) Sec. 4.14 ;

J. Schwinger, Proc. Natl. Acad. Sci. USA, {\bf 47} 570 (1960);   
Quantum Kinematics and Dynamics, (W. A. Benjamin Inc., New York, 1970).
\bibitem{balian} Balian R and Itzykson C 1986 {\it C. R. Acad. Sci. Paris} {\bf 303} 773.
\bibitem{zachos}  D. B. Fairlie, et. al, Phys. Lett. B{\bf 218}, 203 (1989)\\
D. B. Fairlie, et. al, Phys. Lett. B{\bf 224}, 101 (1989).
\bibitem{manol} E. G. Floratos, Phys. Lett. B{\bf 228}, 335 (1989);{\it ibid}
B{\bf 233}, 395 (1989),\\
Athanasiu G G et al 1994 {\it Nucl. Phys. B}{\bf 425} 343\\
Athanasiu G G et al 1996 {\it J. Phys. A: Math. Gen.}{\bf 29} 6737.
\bibitem{abe} 
E. Abe, Hopf Algebras, (CUP, Cambridge 1980).
\bibitem{dembost} D. Ellinas, Proceedings of "IV Workshop on Physics and 
Computation, PhysComp96" p. 108 
Eds, T. Toffoli et. al, (New England 
Complex Systems Institute 1996), also in http://pm.bu.edu/PhysComp96/ .
\bibitem{megalo} Ellinas D and Floratos E G to appear.

\end{thebibliography}
\end{document}